\documentclass[aps,pre,twocolumn,superscriptaddress]{revtex4}

\usepackage{graphicx}% Include figure files
\usepackage{dcolumn}% Align table columns on decimal point
\usepackage{bm}% bold math
\usepackage{color} 
\usepackage{ulem}

\newcommand{\xmg}[1]{#1}

\begin{document}

%Title of paper
\title{Jamming of Frictional Particles: a Non-equilibrium First Order Phase Transition}

\author{Matthias Grob}
 \affiliation{Max Planck Institute for Dynamics and
   Self-Organization, Am Fa\ss berg 17, 37073 G\"ottingen, Germany}

 \author{Claus Heussinger}\affiliation{Institute for Theoretical
   Physics, Georg-August University of G\"ottingen, Friedrich-Hund
   Platz 1, 37077 G\"ottingen, Germany}

\author{Annette Zippelius}\affiliation{Max Planck Institute for
  Dynamics and Self-Organization, Am Fa\ss berg 17, 37073 G\"ottingen,
  Germany}\affiliation{Institute for Theoretical Physics,
  Georg-August University of G\"ottingen, Friedrich-Hund Platz 1,
  37077 G\"ottingen, Germany}

\date{\today}

\begin{abstract}
 We propose a phase diagram for the shear flow of dry granular
   particles in two dimensions based on simulations and a
   phenomenological Landau-theory for a nonequilibrium first order
   phase transition. Our approach incorporates both frictional as well
   as frictionless particles. The most important feature of the
   frictional phase diagram is re-entrant flow and a critical jamming
   point at finite stress. In the frictionless limit the regime of
   re-entrance vanishes and the jamming transition is continuous with
   a critical point at zero stress.  The jamming phase diagrams
   derived from the model agree with the experiments of Bi {\it et
     al.} (Nature (2011)) and brings together previously conflicting
   numerical results.
% The flow curves, i.e. stress $\sigma$ versus strain rate
%$\dot{\gamma}$, show hysteresis above a critical packing fraction
%$\varphi_c=0.7922$, jumping between an inertial flow phase with small $\sigma$
%to a plastic flow phase with high $\sigma$. A yield stress is observed for
%$\varphi>\varphi_{\sigma}=0.8003$ and the viscosity diverges at
%$\varphi_{\eta}=0.818\xmg{5}$.  The transition from flowing to jammed states is
%re-entrant with transient jam-and-flow states in between. All these features can
%be explained by the simple model: The viscosity diverges, when the inertial flow
%regime becomes locally unstable and the transient jam-and-flow states are
%interpreted as metastable states.
 \end{abstract}


\maketitle

Random-close packing is the point at which hard-spherical -- and frictionless --
particles generally jam into a stable heap. It is now known that the precise
close-packing density $\varphi_{\rm rcp}$ depends on the preparation
protocol~\cite{PhysRevLett.104.165701}.  Nevertheless, this variability is small
when compared to frictional systems, i.e. systems where particles not only
transmit normal forces but also tangential forces among themselves. Indeed,
frictional systems can jam at densities anywhere between random-close,
random-loose or even random-very-loose
packing~\cite{ciamarra12:_statis, song2008Nature}.
In this paper we deal with the \textit{flow} properties of frictional
granular systems, where the jamming transition can be studied by
monitoring the flow curves, i.e. the stress-strain rate relations
$\sigma(\dot\gamma)$. Previous simulations performed in the
hard-particle limit~\cite{cruz2005PRE,PhysRevLett.109.118305} do not
observe any qualitative difference between frictionless and frictional
systems, other than a mere shift of the critical density from
$\varphi_{\rm rcp}$ to $\varphi_J(\mu)$, which depends on the friction
coefficient $\mu$ of the particles. Similar results, accounting for
particle stiffness, are presented in
Refs.~\cite{PhysRevE.85.021305,PhysRevE.60.6890}. Quite in contrast,
Otsuki { \it et al.}~\cite{otsuki2011critical} recently observed a
discontinuous jump in the flow curves of the frictional system, which
is absent in the frictionless
analogue~\cite{otsuki2009critical}. In addition, they find not one
  but three characteristic densities for the jamming transition, which
  degenerate into random-close packing when $\mu\to 0$.
  Similarly, Ciamarra {\it et al.}~\cite{ciamarra2011jamming} observe
  three (but different) jamming transitions.
%not one but three jamming transition lines that degenerate into random-close
%packing when $\mu\to 0$.
Experimentally, Bi {\it et al.}  ~\cite{bi2011jamming} present a jamming phase
diagram with a non-trivial (re-entrance) topology that is not present in the
frictionless scenario.

These latter results hint at friction being a non-trivial and indeed
``relevant'' perturbation to the jamming behavior of granular
particles.  Unfortunately, several inconsistencies remain
  unresolved. For example, the phase diagram in
\cite{ciamarra2011jamming} is different from \cite{bi2011jamming} and
does not show stress jumps as observed in
\cite{otsuki2011critical}. This points towards a more fundamental lack
of understanding of the specific role of friction in these
systems. What is the difference between frictional and frictionless
jamming? By
  combining mathematical modelling with strain- and stress-controlled
  simulations we propose a jamming scenario that not only encompasses
  frictional as well as frictionless systems, but also allows to bring
  together previously conflicting results.

We simulate a two-dimensional system of $N=8000$ soft, frictional
particles in a square box of linear dimension $L$. The particles all
have the same mass $m=1$, but are polydisperse in size: $2000$
particles each for diameter $d=0.7, 0.8, 0.9, 1.0$.  The particle
volume fraction is defined as $\varphi = \sum_{i=1}^N \pi
d_i^2/4L^2$. Normal and tangential forces, $\bm{f}^{(n)}$ and
$\bm{f}^{(t)}$, are modeled with linear springs of unit strength for
both, elastic as well as viscous contributions. (Thereby units of
time, length and mass have been fixed). Coulomb friction is
implemented with friction parameter
$\mu=2$~\xmg{\footnote{\xmg{The variation of $\mu$ has been studied by 
Otsuki { \it et al. }, see Figs. 9 and 15 and the related discussion in 
\cite{otsuki2011critical}. Near $\mu=2$ the rheological properties depend weakly on 
this parameter and $\mu=2$ can be regarded as the limit of large friction.}}}. In the strain-controlled
simulations, we prepare the system with a velocity profile
$\bm{v}_{flow} = \dot{\gamma}(0) y \bm{\hat{e}}_x$
initially. Subsequently the shear rate is implemented with
Lees-Edwards boundary conditions~\cite{lees1972computer} until a total
strain of $200\%$ is achieved after time $T$. Whenever the strain rate
is changed to a new value, we wait for a time $\sim 0.5T$ to allow for
the decay of transients. In the stress-controlled simulations, a
boundary layer of particles is frozen and the boundary at the top is
moved with a force $\sigma L\bm{\hat{e}}_x $, whereas the bottom plate
remains at rest.

\begin{figure}
  \includegraphics[angle=0, width=1.0\linewidth]{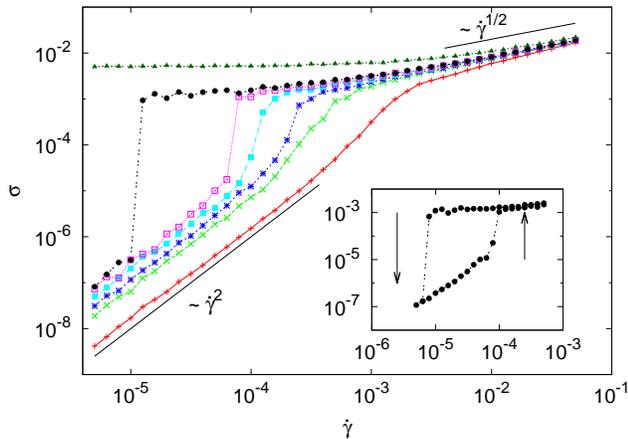}
  \caption{\label{fig:flowcurves}Flow curves $\sigma(\dot{\gamma})$
    for different packing fractions $\varphi = 0.78, 0.7925, 0.795,
    0.7975, 0.79875, 0.80, 0.82$ (from bottom to top). Main part: flow
    curves obtained by decreasing $\dot{\gamma}$; Inset: example of a
    hysteresis loop for $\varphi=0.80$.}
\end{figure}

In the strain-controlled simulations we impose the strain rate
$\dot{\gamma}$ and measure the response, the shear stress
$\sigma(\dot{\gamma})$, for a range of packing fractions
$0.78\leq\varphi\leq 0.82$. Thereby the system is forced to flow for all
packing fractions; the resulting flow curves are shown in
Fig.~\ref{fig:flowcurves}.

We observe three different regimes.  For low packing fraction, the
system shows a smooth crossover from Bagnold scaling,
$\sigma=\eta\dot{\gamma}^2$ (called ``inertial flow'') to
$\sigma\propto\dot{\gamma}^{1/2}$ (called ``plastic flow''). As
the packing fraction is increased, we observe a transition to
hysteretic behaviour~\cite{otsuki2011critical}: decreasing the strain
rate from high values, the system jumps discontinuously to the lower
branch. Similarly, increasing the strain rate from low values, a jump
to the upper branch is observed. 
A well developed hysteresis loop is
shown in the inset of Fig.~\ref{fig:flowcurves}.  
\xmg{The onset of hysteresis defines the critical density $\varphi_c$. We estimate 
its value, $\varphi_c$, between $0.7925$ and $0.795$
by visual inspection of the flow curves as described in the supplemental material~\cite{supplement}.
%linear fit of the magnitude of the stress discontinuity, i.e. $\Delta \sigma \propto \varphi - \varphi_c$.
As $\varphi$ is increased beyond the critical value $\varphi_c$, the jump
to the lower branch happens at smaller and smaller $\dot{\gamma}$,
until at $\varphi_{\sigma}$, the upper branch first extends
to zero strain rate, implying the existence of a yield stress,
$\sigma_{yield}$. For $\varphi_c<\varphi< \varphi_{\sigma}$, the strain rate
for the jump to the lower branch, $\dot{\gamma}_{\sigma} \propto \varphi_{\sigma} - \varphi$, scales linearly with the distance to $\varphi_{\sigma}$
which allows us to determine $\varphi_{\sigma} \cong 0.8003$. 
Finally at $\varphi_{\eta}$, the generalized viscosity,
$\eta=\sigma/\dot{\gamma}^2$, diverges and for
$\varphi>\varphi_{\eta}$ only plastic flow is observed. The scaling of
the viscosity $\eta \propto (\varphi_{\eta}-\varphi)^{-4}$ is in agreement with previous
results~\cite{otsuki2011critical} and yields $\phi_{\eta}\cong0.819$. Note, that all three 
packing fractions are well seperated and furthermore $\varphi_\eta$ is still
well below the frictionless jamming density at random close packing
$\varphi_{\rm rcp}\cong 0.8433$. The scaling plots $\dot{\gamma}_{\sigma}$ 
and $\eta$ are shown in the supplemental material~\cite{supplement}.}

%\begin{figure}
%   \includegraphics[width=1.0\linewidth]{FitViscAndDrop}
%   \caption{\label{fig:FitVisc} Left: Fit of the strain rate
%     $\dot{\gamma}_{\sigma}$, where the plastic stress in the upper
%     branch drops to zero to $\dot{\gamma}_{\sigma} \propto
%     (\varphi_{\sigma}-\varphi)^{\alpha}$, yielding $\varphi_{\sigma} =
%     0.8003(2)$ and $\alpha = 1.0(1)$. Right: Fit of the inverse
%     generalized viscosity to $\eta^{-1} \propto (\varphi_{\eta}-\varphi)^4$
%     yielding $\varphi_{\eta}=0.819(1)$.}
%\end{figure}

All the observations can be explained in the framework of a simple model, 
  which can be viewed as a phenomenological Landau theory, that interpolates
smoothly between the inertial and the plastic flow regime:
%\xx{maybe set $b\to -b$ so we can define $b$ positive and the
%  frictionless limit corresponds to $b\to 0$.}
\begin{equation}
  \label{eq:fc1}
  \dot{\gamma} (\sigma) = a\sigma^{1/2} - b\sigma + c\sigma^2,
\end{equation}
where $a,b,c$ are coefficients which in general depend on packing fraction.
Eq.~(\ref{eq:fc1}) can be taken to result from a class of constitutive models
that combine hydrodynamic conservation laws with a microstructural evolution
equation~\cite{olmsted2008perspectives}, or from mode-coupling
approaches~\cite{holmes05:_glass}.

The numerical data suggest that the plastic flow regime is only
weakly density dependent \xmg{for packing fractions considered here}, so we take $c$ to be independent of
$\varphi$ for simplicity. In the inertial flow regime, on the other
hand, we expect to see a divergence of the shear viscosity at
$\varphi_{\eta}$, implying that the coefficient $a$ of our model
vanishes at $\varphi_\eta$ and changes sign, $a=a(\varphi)=a_0
|\varphi_{\eta} - \varphi| (\varphi_{\eta} - \varphi)$. The
coefficient $b$ is assumed to be at most weakly density dependent.

\begin{figure}
  \centering
  \includegraphics[width=1.0\linewidth]{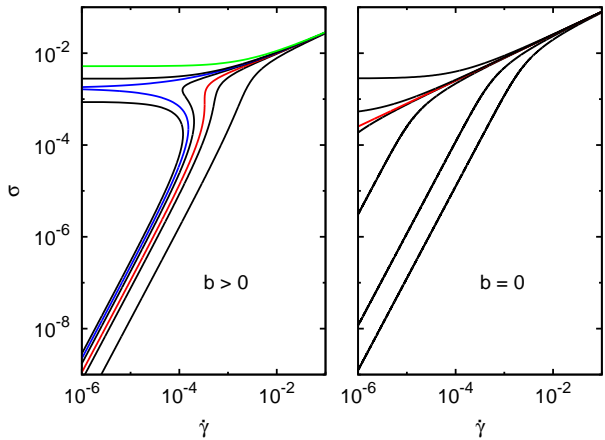}
  \caption{\label{fig:schematic} \xmg{Flow curves of the simple model Eq.~(\ref{eq:fc1}). Left: Frictional scenario with range of packing fractions as in Fig.~\ref{fig:flowcurves}; $\phi_c$ is indicated by the red line, $\phi_{\sigma}$ by the blue line and $\phi_{\eta}$ by the green line. Right: Flow curves for frictionless particles of the simple model implemented with $b = 0$ and $\varphi_{c} = \varphi_\sigma = \varphi_\eta=0.8433$; critical flow curve in red.}} 
\end{figure}

The simple model predicts a discontinuous phase transition with a
critical point in analogy to the van der Waals theory of the
liquid-gas transition (see Fig.~\ref{fig:schematic}a). The critical
point is determined by locating a vertical inflection point in the
flow curve. In other words we require $\partial_{\sigma}\dot{\gamma} =
0$ and simultaneously $\partial_{\sigma\sigma}\dot{\gamma} = 0$. These
two equations together with the constitutive equation (\ref{eq:fc1})
determine the critical point: $b_c =
\frac{3}{2}a(\varphi_c)^{2/3}c^{1/3}$ with the critical strain rate
given by $\dot{\gamma}_c = \frac{3}{16}\frac{a^{4/3}}{c^{1/3}}$ and
the critical stress $\sigma_c= \frac{1}{4} (\frac{a}{c})^{2/3}$. For
$\varphi>\varphi_c$, the model predicts an unstable region, where
$\partial_{\sigma}\dot{\gamma} < 0$. This is where the stress jump
occurs in the simulations. The flow curves of the model are presented
in Fig.~\ref{fig:schematic}a, assuming $b \equiv b_c$, and
fitting the two constants $c,a_0$ to the data.  The model predicts a
yield stress to first occur, when two (positive) zeros for the
function $ \dot{\gamma} (\sigma)=0$ coincide. This happens at a
density $\varphi_{\sigma}$ determined implicitly by
$a(\varphi_c)=a(\varphi_{\sigma})\sqrt{2}$ and the yield stress is
given by $\sigma_{yield} = (a/(2c))^{2/3}$. The flow curves can be
fitted better, if we allow for weakly density dependent coefficients
$b$ and $c$. However, we refrain from such a fit, because even in its
simplest form the model can account for all observed features
qualitatively: a critical point at $\varphi_c$, the appearance of a
yield stress at $\varphi_{\sigma}$ and the divergence of the viscosity
at $\varphi_{\eta}$, ordered such that
$\varphi_c<\varphi_{\sigma}<\varphi_{\eta}$. \xmg{The flow curves for these three packing fractions 
are highlighted in Fig.~\ref{fig:schematic} and 
further illustrated in the supplemental material~\cite{supplement}.}

%frictional vs frictionless
\xmg{The limiting case of frictionless particles can be reached by
  letting $\mu\to 0$. Simulations indicate that in this limit
  hysteretic effects
  vanish~\cite{otsuki2011critical,otsuki2009critical} and the jamming
  density is increased approaching random close packing. Within the model
  this transition can be understood in terms of the variation of
  two parameters: First $b(\mu)\to 0$ in Eq.~(\ref{eq:fc1}) implies that the three densities $(\varphi_c, \varphi_{\sigma}, \varphi_{\eta})$ coincide and second
  $\varphi_{\eta}(\mu)\to\varphi_{\rm rcp} \cong 0.8433$.} While a $\mu$-dependent
  $\varphi_\eta$ simply shifts the phase diagram towards higher
  densities, the parameter $b$ accounts for the more important changes
  of the topology of the phase diagram. The flowcurves in this limit
  are presented in Fig~\ref{fig:schematic}b. They present a continuous
  jamming scenario consistent with previous simulations in
  inertial~\cite{otsuki2009critical} as well as overdamped
  systems~\cite{PhysRevLett.109.108001,PhysRevLett.105.088303}.

What happens in the unstable region? Naively one might expect
``coexistence'' of the inertial and the plastic flow regime,
i.e. shear banding. However, this would have to happen along the
vorticity direction~\cite{dhont2008gradient,olmsted2008perspectives},
which is absent in our two-dimensional setting.  Alternative
possibilities range from oscillating to chaotic
solutions~\cite{PhysRevE.66.025202,nakanishi2012fluid}.  We will see
that, instead, the system stops flowing and jams at intermediate
stress levels. Interestingly, this implies re-entrance in the
$(\sigma,\varphi)$ plane with a flowing state both, for large and small
stress, and a jammed state in between.

\begin{figure}
  \centering
  \includegraphics[width=1.0\linewidth]{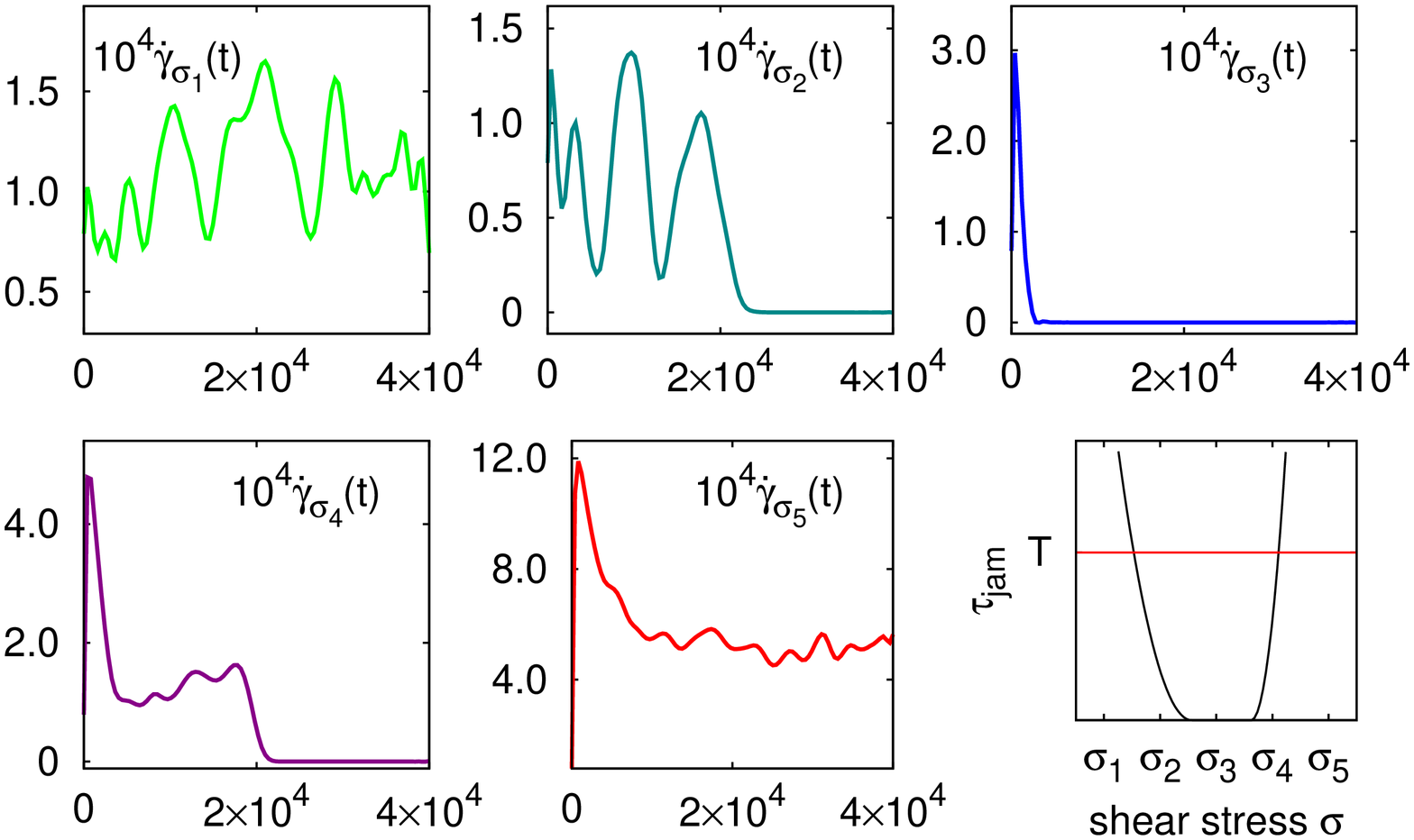}
  \caption{\label{fig:timeseries07975}Time series of the strain rate
    with packing fraction $\varphi=0.7975$ and different, but fixed,
    stresses $\sigma_1 < \sigma_2 < \sigma_3 < \sigma_4 < \sigma_5$
    (stress values are indicated in Fig.~\ref{fig:full}). Lower
    right corner: Schematic picture of the jamming time being cut off
    at the simulation time $T$.}
\end{figure}

To address the unstable regime in more detail, we have performed
stress-controlled simulations: the shear stress is imposed and we
measure the strain rate as a function of time. The initial
configuration\xmg{s are} chosen with a flow profile corresponding to the
largest strain rate in the inertial flow regime, which was observed
previously in the strain-controlled experiments. \xmg{For a fixed $\varphi$,} several time series
are shown in Fig.~\ref{fig:timeseries07975}, representing the
different 
%\xmg{flow}
regimes. The lowest value of $\sigma_1$ is chosen in
the inertial flow regime, so that the system continues to flow for
large times. Similarly $\sigma_5$ is chosen in the plastic flow regime
and the system continues to flow as well. The intermediate value
$\sigma_3$ is chosen in the unstable region and the system immediately
jams.  Between the jamming and the flow regime we find intermediate
phases with transient flow that ultimately
stops~\cite{ciamarra2011jamming}. 

\xmg{To quantify the different flow
regimes, we introduce the time $\tau_{jam}$ the system needs to
jam. Schematically we expect the result shown in the lower right
corner of Fig.~\ref{fig:timeseries07975}: In the jammed phase
$\tau_{jam}=0$, whereas in the flow phase $\tau_{jam}$ is infinite. In between $\tau_{jam}$ is
  finite implying transient flow before the system jams. We expect $\tau_{jam}$ to go to zero as the jammed phase is
approached and to diverge as the flow phases are approached. Given
that the simulation is run for a finite time, the divergence should be cut
off at the time of the simulation run, $T$, indicated by the horizontal (red) line in the schematic in the lower right
corner of Fig.~\ref{fig:timeseries07975}.}

\begin{figure}
  \centering
  \includegraphics[width=1.0\linewidth]{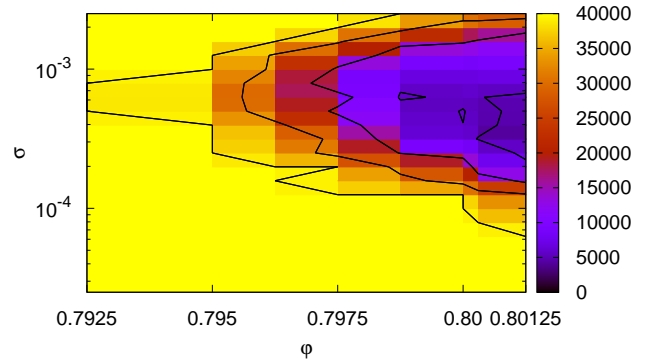}  
  \caption{\label{fig:phases}Numerical results for the phase diagram. The mean
    flow time $\tau_{jam}$ is encoded with color.  The flow phase is
      indicated
%$\tau_{jam} > T = 40000$ 
in yellow (bright) and the jamming phase in blue (dark). Lines are
contours of constant $\tau_{jam}$.}
\end{figure}

%\xmg{I.e. $\tau_{jam}$ depends systematically on $\sigma$ and $\varphi$.}
%For every considered packing fraction and shear stress a set of initial configurations were used to get representative data. 
These expectations are borne out by the simulations: In
Fig.~\ref{fig:phases} we show a contour plot of $\tau_{jam}$ as a
function of $\varphi$ and $\sigma$. In the dark blue region,
$\tau_{jam}$ is very small, corresponding to the jammed state.
%we have $\tau_{jam}\leq\tau_{min}$, corresponding to the jammed state.
 In the bright yellow region $\tau_{jam}$ exceeds the simulation time;
 hence this region is identified with the flow regime -- inertial flow
 for small $\sigma$ and plastic flow for large $\sigma$.  The
 intermediate (red) part of the figure corresponds to the transient
 flow regimes.

\begin{figure}
  \centering
  \includegraphics[width=1.0\linewidth]{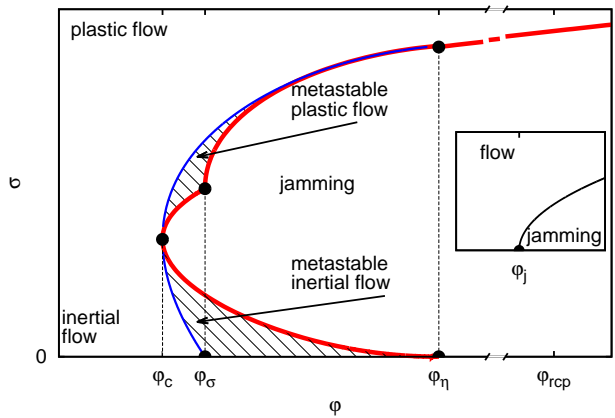}  
  \caption{\label{fig:model_phases}\xmg{Phase diagram of the model
    (schematic), revealing re-entrant flow for small and large
    $\sigma$, as well as flow and jam states in the ``metastable''
    regions for frictional particles (main panel) and the known 
    jamming phase diagram for frictionless particles (inset).}}
\end{figure}

In our simple model Eq.~(\ref{eq:fc1}), the jammed state has to be
identified with the unstable region. It seems furthermore suggestive
to identify the transient flow regime with metastable regions. The
phase diagram, as predicted by the simple model (with finite $b$) is
shown in Fig.~\ref{fig:model_phases} (schematic). In the region
within the (thick) red curve, Eq.(\ref{eq:fc1}) has no solution: the
system jams. Outside the (thin) blue curve, the solution is unique
corresponding to either inertial flow (low stress) or plastic flow
(high stress). In between, in the shaded region, the equations allow
for two solutions and hence metastable states. Jamming from these
metastable states is discontinuous, i.e. the strain rate jumps to zero
from a finite value. At a packing fraction $\varphi_{\sigma}$, a yield
stress first appears and grows as $\varphi$ is increased further,
giving rise to a kink in the red curve and a continuous jamming
scenario. Beyond $\varphi_{\eta}$ inertial flow is no longer
possible~\footnote{The actual transition (``binodal'') has to lie
  somewhere in the shaded region. Its location cannot be constructed
  within the current model (see \cite{dhont2008gradient}).}.
In the frictionless limit $b\to0$ all these different packing
  fractions merge with $\varphi_{\rm rcp}$ giving the phase diagram
  the simple structure well known from previous
  work~\cite{heussingerPRL2009} \xmg{and shown in the inset in 
  Fig.~\ref{fig:model_phases}.}

\begin{figure}
  \centering
  \includegraphics[width=1.0\linewidth]{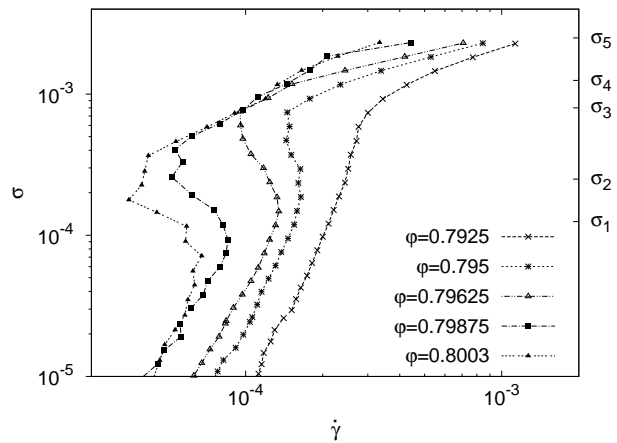}
  \caption{\label{fig:full}Flow curves from the stress-controlled simulations.
    The unstable branches (decreasing stress) are obtained as time averages over
    the transient flow (right axis: stress values used for the time series
      in Fig.~\ref{fig:timeseries07975}).
%% \xmg{COMMENT: are we going to change
%%       something with this figure? we could put the description into the caption
%%       and change the figure's aspect ratio to save some space.  or, in a few
%%       days we could add numerical results for flow curves in the frictionless
%%       setting (they are running right now and could be compared to the model
%%       with $b_c \equiv 0$.)}  
}
\end{figure}

The presence of long transients is fully consistent with the results
of Ref.~\cite{ciamarra2011jamming}. Due to a restricted stress range
in those simulations, however, only the upper part of the phase
diagram is captured and the re-entrance behavior is missed. To get a
better understanding of these transients (or possibly metastable
states), we have tried to construct the flow curves in this regime by
the following procedure: The monitored time series are truncated as
soon as the system jams. The (transiently) flowing part of the
time series is averaged over time, giving rise to the flow curves,
shown in Fig.~\ref{fig:full}. These flow curves show clearly a
non-unique relation $\sigma(\dot{\gamma})$ or equivalently a
non-\xmg{monotonic} relation $\dot{\gamma}(\sigma)$, which can only be
observed as transient behavior, before the system has settled into a
stationary state.

%conclusion:
 In conclusion: the goal of this paper is to understand the role
  of friction in the jamming behavior of dry granular matter. To this
  end we present a theoretical model (supplemented by MD simulations)
  that can reproduce all the phenomenology of simulated flowcurves
  (Fig.~\ref{fig:schematic}) both for the fully frictional system as
  well as for the limiting case of frictionless particles. The jamming
  phase diagrams derived from the model agree with recent
  experiments~\cite{bi2011jamming}. The key result is that the
  transition between the two jamming scenarios,
  frictionless/continuous and frictional/discontinuous, can in our
  model be accounted for by the variation of just a single parameter
  ($b$). The most important feature of the frictional phase diagram is
  re-entrant flow and a critical jamming point at finite stress. The
  fragile ``shear jammed'' states observed in the
  experiments~\cite{bi2011jamming} then correspond to the re-entrant
  (inertial) flow regime in our theory. Our work allows to bring
  together previously conflicting
  results~\cite{PhysRevE.60.6890,PhysRevE.85.021305,ciamarra2011jamming,otsuki2011critical}
  and opens a new path towards a theoretical understanding of a
  unified jamming transition that encompasses both frictionless as
  well as frictional particles.

\begin{acknowledgments}
  We thank Till Kranz for fruitful discussions. We gratefully acknowledge
  financial support by the DFG via FOR 1394 and the Emmy Noether program (He
  6322/1-1).
\end{acknowledgments}

%\bibliography{references.all,en}

\end{document}